\documentstyle[12pt]{article}
\def\cN{{\cal{N}}}
\def\cNtot{{\cal{N}_{\rm tot}}}
\def\cF{{\cal F}}
\def\cNf{\cN_{\mbox{\tiny FS}}}
\def\cNi{\cN_{\mbox{\tiny IS}}}
\def\cNif{\cN_{\mbox{\tiny I/F}}}
\def\cNll{\cNtot^{(ll)}}
\def\cNql{\cNtot^{(ql)}}
\def\cNqq{\cNtot^{(qq)}}

\def\cR{{\cal{R}}}

\def\t12{\Theta_{12}}
\def\akikii{\widehat{k_1k_2}}
\def\akiki{\widehat{k_1k_1}}
\def\akiikii{\widehat{k_2k_2}}
\def\aqiqii{\widehat{q_1q_2}}
\def\akiqii{\widehat{k_1q_2}}
\def\akiqi{\widehat{k_1q_1}}
\def\akiiqi{\widehat{k_2q_1}}
\def\akiiqii{\widehat{k_2q_2}}
\def\aqiqi{\widehat{q_1q_1}}
\def\aqiiqii{\widehat{q_2q_2}}
\def\akipi{\widehat{k_1p_1}}
\def\aqipi{\widehat{q_1p_1}}
\def\akipii{\widehat{k_1p_2}}
\def\aqipii{\widehat{q_1p_2}}
\def\akiipi{\widehat{k_2p_1}}
\def\aqiipi{\widehat{q_2p_1}}
\def\akiipii{\widehat{k_2p_2}}
\def\aqiipii{\widehat{q_2p_2}}
\def\apipii{\widehat{p_1p_2}}
\def\akiri{\widehat{k_1r_1}}
\def\aqiri{\widehat{q_1r_1}}
\def\akirii{\widehat{k_1r_2}}
\def\aqirii{\widehat{q_1r_2}}
\def\akiiri{\widehat{k_2r_1}}
\def\aqiiri{\widehat{q_2r_1}}
\def\akiirii{\widehat{k_2r_2}}
\def\aqiirii{\widehat{q_2r_2}}
\def\aririi{\widehat{r_1r_2}}
\def\apirii{\widehat{p_1r_2}}
\def\ariri{\widehat{r_1r_1}}
\def\apipi{\widehat{p_1p_1}}
\def\apiipii{\widehat{p_2p_2}}
\def\ariirii{\widehat{r_2r_2}}
\def\apiiri{\widehat{p_2r_1}}
\def\apiirii{\widehat{p_2r_2}}
\def\apiri{\widehat{p_1r_1}}

\def\beq{\begin{equation}}   \def\eeq{\end{equation}}
\def\beeq{\begin{eqnarray}}  \def\eeeq{\end{eqnarray}}
\def\lapp{{\ \lower 0.6ex \hbox{$\buildrel<\over\sim$}\ }}
\def\gapp{{\ \lower 0.6ex \hbox{$\buildrel>\over\sim$}\ }}
\newcommand{\gapproxeq}{\lower .7ex\hbox{$\;\stackrel{\textstyle >}{\sim}\;$}}
\newcommand{\lapproxeq}{\lower .7ex\hbox{$\;\stackrel{\textstyle <}{\sim}\;$}}
\def\to{\rightarrow}

\def\qq{q\bar{q}}
\def\ff{f\bar{f}}

\def\fbar{{\bar{f}}}
\def\tt{t\bar{t}}
\def\ww{W^+W^-}

\def\ffff{f\bar{f}f'\bar{f'}}
\def\degree{^{\circ}}

\def\ee{e^+e^-}

\def\mw{M_W}
\def\gw{\Gamma_W}

\def\tdkbis{1}
\def\lepcc{2}
\def\dkos{3}
\def\kos{4}
\def\jikia{5}
\def\veltman{6}
\def\book{7}
\def\jpsi{8}

\begin{document}
\begin{titlepage}
\vspace*{-1cm}
\begin{flushright}
 DTP/93/16   \\
 UCD-93-11 \\
 LU-TP-93-9 \\
 hep-ph/9305327 \\
 May 1993 \\
\end{flushright}
\vskip 1.cm
\begin{center}
{\Large\bf Soft Photons  in $\ww$ Production at LEP200}
\vskip 1.cm
{\large Yu.L. Dokshitzer}
\vskip .2cm
{\it Department of Theoretical Physics, University of Lund \\
S\"olvegatan 14A, S-22362 Lund, Sweden }\\
\vskip .4cm
{\large V.A. Khoze}
\vskip .2cm
{\it Department of Physics, University of Durham \\
Durham DH1 3LE, England }\\
\vskip .4cm
{\large Lynne H. Orr\footnote{Address after September 1, 1993:  Department of
Physics and Astronomy, University of Rochester, Rochester, NY 14627, USA}}
\vskip .2cm
{\it Department of Physics, University of California \\
Davis, CA 95616, USA. } \\
\vskip .4cm
and
\vskip   .4cm
{\large  W.J. Stirling}
\vskip .2cm
{\it Departments of Physics and Mathematical Sciences, University of Durham \\
Durham DH1 3LE, England }\\
\vskip 1cm
\end{center}
\begin{abstract}
The pattern of soft photon radiation in $\ee\to\ww$ has a rich
structure, with contributions from photon emission off the initial state
and off  the final state particles both before and after decay.
In particular, the
interference between the contributions
 involving the decaying $W$'s depends on the decay width.
 We review the theoretical result for the
radiation pattern, and present predictions for LEP200, {\it i.e.}
in $\ee$ annihilation just above $\ww$ threshold.
\end{abstract}
\vfill
\end{titlepage}
\newpage
%%%%%%%%%%%%%%%%%%%%%%%%%%%%%%%%%%%%%%%%%%%%%%%%%%%%%%%%%%%%%%%%%%%%%%
\section{Radiation Pattern Near Threshold}

Heavy unstable charged particles such as the $W$ boson can emit photons
 before and after
they decay. The relative size of the two contributions, and consequently
the overall radiation pattern, depends
sensitively on the timescale of the emission compared to the lifetime of
the unstable particle [\tdkbis].
A particularly important process which exhibits these effects is
soft photon production in $\ee\to\ww\to\ffff$ at energies just above threshold,
which will be studied at LEP200 [\lepcc].
The radiation pattern of a soft
photon of energy $\omega$ is sensitive to the $W$ decay width for $\omega
\sim\Gamma_W$.
In a previous study we have derived some general results for the radiation
pattern for this process [\dkos], and in this Letter we present
specific numerical predictions for LEP200.

The general formalism for calculating the distribution of soft radiation
in a process involving the production and decay of unstable particles
can be found in references [\dkos--\jikia]. The differential
distribution for the production of a  soft
photon with momentum $(k)$ in the process
\beq
e^-(k_1) + e^+(k_2) \to W^-(q_1) + W^+(q_2) \to f(p_1) + \fbar(r_1)
+ \fbar'(p_2) + f'(r_2)    \; ,
\label{process}
\eeq
is given by
\beq
{1\over N}\ {d N\over d\omega\> d\cos\theta \> d\phi}\ = \
{\alpha\over 4 \pi^2} \ \omega \ \cF \; ,
\eeq
where $\omega$ (the photon energy), $\theta$ and $\phi$ are
 measured in the $\ee$ centre-of-mass frame.
The radiation pattern is described by the function $\cF$. The result for
this is calculated in two steps. First, for the case of soft photon
radiation  in {\it stable} $\ww$ production, we have the well-known
result [\veltman]
\beq
\cF_0  = 2\akikii - \akiki - \akiikii +  2\akiqi -2 \akiqii -2 \akiiqi
+  2\akiiqii + 2  \aqiqii - \aqiqi - \aqiiqii \; ,
\label{general}
\eeq
where the `antennae' are defined by [\book]
\beq
\widehat{pq} \> \equiv \> {p\cdot q \over p\cdot k \ q\cdot k} \; .
\eeq
Note contributions from initial state radiation, final state radiation,
and the interference between them.

For unstable $W$'s, decaying to fermions as in  (\ref{process}),
 we have $\cF_0 \to \cF$ with $\cF$ given
by (\ref{general}) with the replacements:
\begin{eqnarray}
\akikii,\akiki,\akiikii & \to & \akikii,\akiki,\akiikii \nonumber \\
\akiqi & \to & \akiqi +\chi_1\; [Q \akipi + (1-Q)\akiri  -\akiqi ]  \nonumber
\\
\akiqii & \to & \akiqii + \chi_2\; [ Q'\akipii
+ (1-Q') \akirii - \akiqii ]  \nonumber \\
\akiiqi & \to & \akiiqi  + \chi_1\; [ Q\akiipi
+ (1-Q)\akiiri  -\akiiqi ]  \nonumber \\
\akiiqii & \to & \akiiqii + \chi_2\; [Q' \akiipii
+(1-Q') \akiirii -\akiiqii ]  \nonumber \\
\aqiqii & \to & \aqiqii + \chi_2\; [ Q'\aqipii + (1-Q')\aqirii
-\aqiqii ] \nonumber \\
 & &\  +  \chi_1 \; [Q \aqiipi + (1-Q)\aqiiri -\aqiqii ]  \nonumber \\
& & \ + \ \chi_{12} \; [QQ' \apipii -Q'\aqipii -(1-Q')\aqirii
  -Q\aqiipi \nonumber \\
& & - (1-Q)\aqiiri + Q(1-Q')\apirii + Q'(1-Q)\apiiri
 \nonumber \\
& & + (1-Q)(1-Q') \aririi+ \aqiqii ]  \nonumber \\
\aqiqi & \to & 2 \aqiqi  + Q^2 \apipi + (1-Q)^2 \ariri
\nonumber \\
& & +2Q(1-Q)\apiri -2Q\aqipi -2(1-Q)\aqiri  \nonumber \\
 & & + 2 \chi_1 \; [Q \aqipi
+(1-Q)\aqiri -\aqiqi ]  \nonumber \\
\aqiiqii & \to & 2 \aqiiqii +  Q'^2 \apiipii + (1-Q')^2 \ariirii \nonumber \\
& & + 2 Q'(1-Q') \apiirii -2 Q' \aqiipii -2(1-Q') \aqiirii \nonumber \\
& & + 2 \chi_2\;  [Q' \aqiipi
+(1-Q') \aqiirii -\aqiiqii ] \; ,
\label{reps}
\end{eqnarray}
with the charge factors given by $Q=|Q_f|$ and $Q' = |Q_{\bar{f}'}|$,
{\it e.g.} $2/3$ and $1$ for  hadronic and leptonic decays respectively.
Now we have contributions from additional antennae associated
with the decays $W\to f\fbar$, together with contributions
from the interference of photons radiated at the production and decay
stages. This interference is controlled by the  `profile functions'
[\kos,\jikia]
\begin{eqnarray}
\label{chione}
\chi_i & = & {\mw^2\gw^2 \over (q_i\cdot k)^2 + \mw^2 \gw^2 } \\
\label{chitwo}
\chi_{12} & = & {\mw^2\gw^2\; (q_1\cdot k\; q_2\cdot k + \mw^2\gw^2)
 \over ( (q_1\cdot k)^2 + \mw^2 \gw^2)
\; ( (q_2\cdot k)^2 + \mw^2 \gw^2) } \; ,
\end{eqnarray}
 which depend on the $W$ mass ($\mw$) and decay width ($\gw$).
They have the (formal) property that $\chi_i, \chi_{12} \to 0$
as $\gw \to 0$, and $\chi_i, \chi_{12} \to 1$
as $\gw \to \infty$.
Note that only soft photons with energy $\omega \lapproxeq \gw$
can lead to significant interference contributions: the emission
of energetic photons (either real or virtual) with $\omega \gg \gw$
pushes the $W$ propagators far  off their resonant energy and
the interference becomes negligible. This is a well-understood
phenomenon, dating back (at least) to the early days of $J/\psi$
physics\footnote{VAK thanks V.S. Fadin for reminding him of this.}
[\jpsi].

The rather complicated structure implied by (\ref{general})-(\ref{chitwo})
is greatly simplified if we take the `threshold limit' appropriate to LEP 200,
{\it i.e.} $\sqrt{s} \sim 2 M_W$ so that $v_W \ll 1$.
In this case, the radiation off the almost stationary $W$ bosons
is suppressed [\kos], and the only contributions which survive
are radiation off the initial state leptons, radiation off the
two  final state $f \bar f$ antennae, and interference between them. Noting
also
that near threshold
\beq
\chi_1,\; \chi_2,\; \chi_{12} \ \to \chi(\omega) =
{\gw^2 \over \omega^2 + \gw^2} \; ,
\eeq
we find [\dkos]
\beq
\cF \> = \>  {1\over \omega^2} \> \cNtot
\> \equiv \>  {1\over \omega^2} \> (\cNi + \cNf + \cNif )
\label{fsplit}
\eeq
where
\beeq
\label{ni}
\cNi & = & {4 \over \sin^2\theta_0} \\
\label{nf}
%\cNf  & = & (1- \chi(\omega) )\; \left[ Q\;
%{ 1+\cos\theta_1 \over 1-\cos\theta_1} + (1-Q)\;
%{ 1-\cos\theta_1 \over 1+\cos\theta_1} \right. \nonumber   \\
%  &  &\left. +  Q'\; { 1+\cos\theta_2 \over 1-\cos\theta_2} + (1-Q')\;
%{ 1-\cos\theta_2 \over 1+\cos\theta_2}\right]
%-{4Q(1-Q) \over \sin^2\theta_1} -{4Q'(1-Q') \over \sin^2\theta_2} \nonumber \\
%& & + 2 \chi(\omega) \; \left[
%QQ'\; {1-\cos\Theta_{12} \over (1-\cos\theta_1)(1-\cos\theta_2)}
%\right. \nonumber \\
%& & + Q(1-Q')\; {1+\cos\Theta_{12} \over (1-\cos\theta_1)(1+\cos\theta_2)}
%  \nonumber \\
%& & + \; \
%(1-Q)Q'\; {1+\cos\Theta_{12} \over (1+\cos\theta_1)(1-\cos\theta_2)}
%\nonumber \\
%& & \left.
%+ (1-Q)(1-Q')\; {1-\cos\Theta_{12} \over (1+\cos\theta_1)(1+\cos\theta_2)}
% \right]  \\
\cNf  & = &  {[(2Q-1)+\cos\theta_1]^2 \over \sin^2\theta_1}
           + {[(2Q'-1)+\cos\theta_2]^2 \over \sin^2\theta_2}
\; - \; 2 \chi(\omega) \nonumber \\
& & \times\; (\cos\Theta_{12} -\cos\theta_1\cos\theta_2)
\; {[(2Q-1)+\cos\theta_1][(2Q'-1)+\cos\theta_2] \over
\sin^2\theta_1\sin^2\theta_2}  \\
\label{nif}
%\cNif & = & 2\chi(\omega)\;
%\left[ Q\; {1-\cos\Theta_1 \over (1-\cos\theta_0)
%(1-\cos\theta_1) } + (1-Q)\;  {1+\cos\Theta_1 \over (1-\cos\theta_0)
%(1+\cos\theta_1) } \right.  \nonumber \\
%& & - Q'\; {1-\cos\Theta_2 \over (1-\cos\theta_0)
%(1-\cos\theta_2) } - (1-Q')\;  {1+\cos\Theta_2 \over (1-\cos\theta_0)
%(1+\cos\theta_2) }   \nonumber \\
%& & - Q\; {1+\cos\Theta_1 \over (1+\cos\theta_0)
%(1-\cos\theta_1) } - (1-Q)\;  {1-\cos\Theta_1 \over (1+\cos\theta_0)
%(1+\cos\theta_1) }   \nonumber \\
%& & \left. + Q'\; {1+\cos\Theta_2 \over (1+\cos\theta_0)
%(1-\cos\theta_2) } + (1-Q')\;  {1-\cos\Theta_2 \over (1+\cos\theta_0)
%(1+\cos\theta_2) } \right]
\cNif & = & \chi(\omega)\;\cNi\; \left[
(\cos\theta_0\cos\theta_1-\cos\Theta_1)
{(2Q-1)+\cos\theta_1 \over \sin^2\theta_1} \right. \nonumber \\
 & & \left. -  (\cos\theta_0\cos\theta_2-\cos\Theta_2)
{(2Q'-1)+\cos\theta_2 \over \sin^2\theta_2} \right] \; .
\eeeq
The angles are defined as follows (see Fig.~1):
$\theta_1$ is the angle between the photon
and the fermion $f$, $\theta_2$ is the angle between the photon and the
fermion $f'$, $\theta_0$ is the angle between the photon and the incoming
electron $e$, $\Theta_{12}$ is the angle between $f$ and $f'$,
$\Theta_1$ is the angle between $f$ and $e$, and
$\Theta_2$ is the angle between
$f'$ and $e$.
In deriving this above result we have assumed that all fermions are massless;
the corresponding result for massive fermions is given in reference [\dkos].

We see from (\ref{fsplit})-(\ref{nif})  that
the radiation pattern of a soft
photon of energy $\omega$ has a rich structure depending on
the relative orientation of the charged particles, and in particular
is sensitive to the $W$ decay width  through the
profile function $\chi(\omega)$. The sensitivity is evidently largest
for $\omega \sim \gw$.  For larger $\omega $ values $\chi $
becomes small, and the pattern of radiation is simply that of three
independent antennae: the initial $\ee$ antenna and two final state
$\ff$ antennae.  In contrast, for $\omega$ smaller  than $\gw$
we have $\chi \sim 1$, and the interference   between the emission
off the different antennae becomes large (`coherent emission') [\dkos].

\section{Examples}
The remainder of this Letter is devoted to a brief practical study of the
photon
distribution implied by the above results. In particular, we focus on the
role of the $W$ width in determining the shape of the distributions.
Finally, we address the
question of whether the effects we describe are observable,
 and whether the sensitivity to $\gw$ is large enough
to be measurable.

Obviously an important issue is the overall number of events. From
both a theoretical and experimental point of view, the cleanest
final state  $(ll)$ is $l\bar\nu \bar l \nu$, but this also has the smallest
branching ratio.  The event rates for final states involving at least
one hadronic $W$ decay, $(ql)$ and $(qq)$, are of course larger, but here
we  encounter the problem of being unable to distinguish
the quark jet from the
antiquark jet. (We assume that it will always be possible to pair the jets
in  a four-jet event according to which come from the same $W$ decay.
This is certainly true just above  threshold, where the $W$'s decay to
almost  back-to-back  jets.)
We must therefore symmetrize the above result when applied to $\qq$
decays  to allow for this ambiguity.
In addition, with final jets we have additional `secondary' sources
of photons from within the jet (from $\pi^0$ decay {\it etc.}), and so it
will be necessary in practice to isolate the photons from the jet axes.
Some (less stringent) isolation of the photons from the charged leptons
may also be required.

In the present context, the radiation from the initial state electrons
can be regarded as a background, whose effect is minimized by
keeping all final state particles including the photon well away from
the beam direction. We can even imagine an idealized situation
where all the final state particles are transverse to the beam direction,
in which case the initial state contributions are simply
$\cNi = 4 \; , \ \;  \cNif = 0$.
Apart from the overall constant contribution from the initial state radiation,
this situation is very similar to the analysis of soft gluon radiation
in $\tt$ production [\kos,\dkos].

There is, however, one important way we can take advantage of the
initial state radiation.
We are interested in sensitivity to $\chi(\omega)$, and we see from
(\ref{nif}) that the initial--final state interference is proportional
to $\chi(\omega)$.  Furthermore,
we see from (\ref{ni})-(\ref{nif}) that when the final state fermions
are in the transverse plane ($\cos\Theta_1 = \cos\Theta_2 = 0$),
the $\cNif$ contribution is {\it antisymmetric} with respect to the
forward and backward directions, $\cos\theta_0 > 0$ and $<0$
respectively, the other two
contributions being {\it symmetric}.  The forward-backward
asymmetry in this case is then
\beq
\Delta \cNtot  = 2\cNif  =  \chi(\omega)\;
 {8\cos\theta_0 \over \sin^2\theta_0} \;
\left[  {1+(2 Q-1) \cos\theta_1 \over \sin^2\theta_1 }
\; - \;   {1+(2 Q'-1) \cos\theta_2 \over \sin^2\theta_2 } \right]  .
\label{fbasy}
\eeq
The existence of the asymmetry is easy to understand physically.  The
interference is {\it linear} in the initial and final state
charges; the electron and positron pieces correspond to interchanging
the forward and backward directions, and contribute with opposite signs.
Since the asymmetry is directly proportional to $\chi(\omega)$,
it provides, at least in principle,  a method for
extracting $\gw$. This will be illustrated below.

As an aside, we note that in the general case,
$\cNif$ is antisymmetric under the interchanges
$\cos\theta_0 \leftrightarrow -\cos\theta_0$,
$\cos\Theta_{1,2} \leftrightarrow -\cos\Theta_{1,2}$, and therefore
vanishes after integration over the angles between the initial state and
final state antennae (keeping the relative angles between the pairs of
decay products fixed).

The next step, then, is to recast the results of (\ref{ni})-(\ref{nif})
into the appropriate form for the three types of final state.
This is a straightforward process, and the explicit expressions for the
three functions $\cNll$, $\cNql$ and $\cNqq$ are given
in reference [\dkos] (Eqs.~(D.6),(D.7)).
In particular,
the double leptonic decay result $\cNll$ is obtained by setting $Q=Q'=1$.
For one leptonic decay and one hadronic decay ($\cNql$) we set $Q=2/3$, $Q'=1$,
and symmetrize between the quark and antiquark directions, {\it i.e.}
$Q \leftrightarrow 1 - Q$.  This is equivalent to
$\cos\theta_1 \leftrightarrow - \cos\theta_1$,
$\cos\Theta_1 \leftrightarrow - \cos\Theta_1$,
and $\cos\Theta_{12} \leftrightarrow - \cos\Theta_{12}$, or
dropping terms linear in $(2Q-1)$.
Finally, setting $Q=Q'= 2/3$ and symmetrizing in both the
`1' and `2' angles gives     $\cNqq$.

The following examples illustrate
 the effect of $\gw$ on the radiation pattern.
In each case we take the final state fermions to be in the plane transverse
to the beam.  This is done merely for convenience; the effect of the
forward-backward asymmetry discussed above is straightforward to see
in such configurations.

First consider the case of a four-jet final state where all four jets
are in the transverse plane and $\Theta_{12}$ is the (acute) angle
between a jet from each $W$, labelled ``1" and ``2". After the
appropriate symmetrization we have
%\beeq
%\cNqq & = & {4 \over \sin^2\theta_0}
%+  (1- \chi(\omega) )\; \left[
%{ 1+\cos^2\theta_1 \over \sin^2\theta_1} +
%{ 1+\cos^2\theta_2 \over \sin^2\theta_2} \right] \nonumber \\
%& & - {1\over 9} \left[ {1  \over \sin^2\theta_1} + {1 \over \sin^2\theta_2}
%\right]
%+ 2 \chi(\omega) \; \left[ {1-\cos\Theta_{12}\cos\theta_1\cos\theta_2
%\over \sin^2\theta_1 \sin^2\theta_2} \right] .
%\eeeq
\beeq
\cNqq & = & {4 \over \sin^2\theta_0}  +
   { {1\over 9} +\cos^2\theta_1 \over \sin^2\theta_1} +
   { {1\over 9} +\cos^2\theta_2 \over \sin^2\theta_2} \nonumber \\
& & +\;  2 \chi(\omega) \; \left[
{\cos\theta_1\cos\theta_2(\cos\theta_1\cos\theta_2
-\cos\Theta_{12})\over \sin^2\theta_1 \sin^2\theta_2} \right. \nonumber
\\
& & \left.
+ 2 {\cos\theta_0\over \sin^2\theta_0}
\left( { 1 \over \sin^2\theta_1}- { 1 \over  \sin^2\theta_2} \right)\right] .
\eeeq

The second example is for the one hadronic -- one leptonic $W$ decay
final state. With the jets and leptons again in the transverse plane
we have
\beeq
\label{ql}
%\cNql & = & {4 \over \sin^2\theta_0}
%+  (1- \chi(\omega) )\; \left[
%{ 1+\cos^2\theta_1 \over \sin^2\theta_1} +
%{ 1+\cos\theta_2 \over 1-\cos\theta_2} \right]
% - {1\over 9} \left[ {1  \over \sin^2\theta_1}
%\right] \nonumber \\
%& & +\;  2 \chi(\omega) \; \left[ {1-\cos\Theta_{12}\cos\theta_1
%\over \sin^2\theta_1 (1-\cos\theta_2 ) } \right. \nonumber \\
%& & \left. + 2  {\cos\theta_0\over \sin^2\theta_0}
%\left( { 1 \over \sin^2\theta_1}- { 1 \over 1- \cos\theta_2} \right)\right] .
\cNql & = & {4 \over \sin^2\theta_0}  +
   { {1\over 9} +\cos^2\theta_1 \over \sin^2\theta_1} +
   { 1 +\cos\theta_2 \over 1 - \cos\theta_2}\; +\;  2 \chi(\omega) \nonumber \\
& & \times
\left[ {\cos\theta_1(\cos\theta_1\cos\theta_2 - \cos\Theta_{12})
\over \sin^2\theta_1 (1-\cos\theta_2 ) }
+ 2 {\cos\theta_0\over \sin^2\theta_0}
\left( { 1 \over \sin^2\theta_1}- { 1 \over 1- \cos\theta_2} \right)\right] .
\nonumber \\
\eeeq

Finally, the double-leptonic decay, while less interesting from the
practical point of view, shows the most dramatic $\chi$ dependence.
For the transverse configuration described above we have
\beeq
%\cNll & = & {4 \over \sin^2\theta_0}
%+  (1- \chi(\omega) ) \;\left[
%{ 1+\cos\theta_1 \over 1-\cos\theta_1} +
%{ 1+\cos\theta_2 \over 1-\cos\theta_2} \right]
%\; + \; 2 \chi(\omega)  \nonumber \\
%& &\times \left[ {1-\cos\Theta_{12}
%\over  (1-\cos\theta_1 )(1-\cos\theta_2 ) }
% + 2  {\cos\theta_0\over \sin^2\theta_0}
%\left( { 1 \over 1-\cos\theta_1 }- { 1 \over 1- \cos\theta_2} \right)\right]
%,
\cNll & = & {4 \over \sin^2\theta_0}
+  \left[
{ 1+\cos\theta_1 \over 1-\cos\theta_1} +
{ 1+\cos\theta_2 \over 1-\cos\theta_2} \right]
\; + \; 2 \chi(\omega)  \nonumber \\
& &\times  \left[ {\cos\theta_1\cos\theta_2-\cos\Theta_{12}
\over  (1-\cos\theta_1 )(1-\cos\theta_2 ) }
 +   2{\cos\theta_0\over \sin^2\theta_0}
\left( { 1 \over 1-\cos\theta_1 }- { 1 \over 1- \cos\theta_2} \right)\right]
.     \nonumber \\
\eeeq

To illustrate the effects that can arise due to the $W$ width, we show
the photon angular distributions implied by Eq.~(\ref{ql}) for one
particular final state configuration in which the $W^+$ decays leptonically
and the $W^-$ decays to jets.  As stated above, we take the final state
fermions to be in the plane transverse to the beam, and for this example we
choose $\Theta_{12}=90\degree$.  Our angular convention is such that the
electron beam defines the polar axis (so that the leptons and quarks have
$\theta=90\degree$) and the charged lepton direction defines
the photon azimuthal angle  $\phi_0=0\degree$.  Thus
the jet directions are given by $\phi=90\degree$ and $270\degree$.

First we show how the antisymmetry of $\cNif$ gives rise to
a $\chi$-dependent asymmetry in $\cNql$.  Figures 2 (a-c)
show the radiation pattern $\cNql$ as a function of $\theta_0$
for fixed values of $\phi_0$, for $\chi=0$ (solid lines) and
$\chi=1$ (dashed lines).
Fig.~2(a) corresponds to $\phi_0=0\degree$, {\it i.e.}
the photon is in the plane defined by the beam
and the final state lepton, so that $\theta_2 = |\theta_0 - 90^\circ|$.
The singularities at $\theta_0=0\degree, 90\degree$, and $180\degree$
correspond to the incoming electron, final charged lepton, and incoming
positron, respectively.
Notice the non-zero asymmetry about $\theta_0=90\degree$ for $\chi \neq 0$.
In fact for $\chi = 1$ the interference is so large that the
radiation is almost totally suppressed  at  angles close to $50^\circ$.
At the minimum, there is a two orders of magnitude difference between
the $\chi = 0$ and $\chi = 1$ distributions. In practice, such a
configuration could provide  enhanced sensitivity to $\gw$.  As we shall
see, there is sensitivity to $\gw$ in other configurations as well.

In Fig.~2(b) we again show $\cNql$ as a function of $\theta_0$ at fixed
$\phi_0$, this time for $\phi_0=45\degree$, halfway between the lepton and the
closer jet.  Again  we see an asymmetry, but much smaller in size
than in Fig.~2(a).
When we increase $\phi_0$ to $90\degree$ in Fig.~2(c) to correspond to the
jet direction, we see the asymmetry return, but this time with the
opposite sign.  (This explains the small size of the effect in Fig.~2(b).)
The sign of the asymmetry changes because the quarks' charges are opposite in
sign to that of the lepton, and for this value of $\phi$ the $2Q-1$ term
dominates in the interference; see {\it e.g.} Eq.~(\ref{fbasy}).
Note also that the asymmetry here is less pronounced than in Fig.~2(a):
the (absolute value of the) average charge of the
quarks  is less than the charge of the lepton.

The forward--backward asymmetry is not the only manifestation of width
dependence.  As indicated in Eq.~(\ref{nf}), there are $\chi$-dependent
interference terms in the final state radiation contribution as well; in the
case of soft gluon emission in $e^+e^- \to t \bar t$,
these terms constituted {\it all} of the $\chi$ dependence
[\dkos,\kos].  To illustrate these final-state width effects, we show in
Figure 3  distributions in $\phi_0$ for fixed values of $\theta_0$.
Recall that the final state particles are in the $\theta_0=90\degree$
plane.

Fig.~3(a) shows $\cNql$ for $\theta_0=50\degree$,
roughly halfway
between the electron beam and the final charged lepton, again for $\chi=0$
(solid line) and $\chi=1$ (dashed line).  The difference is striking.
The $\chi=0$ curve is nearly featureless except for slight increases
in the beam--lepton plane at $\phi_0=0\degree, 360\degree$.  In contrast, the
interference suppresses the $\chi=1$ distribution in that same plane, but
enhances it at $\phi_0$ values corresponding to the beam--jet planes.
As we increase $\theta$ and move toward the transverse plane containing the
final state particles (Fig.~3(b), $\theta_0=75\degree$),
we see similar effects, although the $\chi=0$ and $1$ distributions are
less distinct.

Fig.~3(c) shows the distribution for $\theta_0=105\degree$, corresponding
to the backward plane. We see that the width effect is reversed from
the forward plane.  Now the $\chi=1$ curve is enhanced in the beam--lepton
plane while the $\chi=0$ curve exhibits  peaks in the beam--jet planes.
 Finally, in Fig.~3(d) we take $\theta=130\degree$ (between the backward
jet and the positron beam) and the $\chi=0$ curve is again featureless
while the interference induces structure in the dashed curve for $\chi=1$.

Note  that although we have  chosen
 the $ql$ case
to illustrate $\gw$ effects, the $ll$ and $qq$ exhibit significant
sensitivity to $\gw$ as well.  In fact the interference effects in the
$ll$ case
are even more dramatic than those shown here because there is no
charge symmetrization.
In the same way, we expect slightly less sensitivity
in the $qq$ case because the symmetrization applies to both $W$ decays.

Of course, a detailed study of the $ll$ case along the lines of the $ql$ case
described above is likely to  be restricted in practice by the lack of events.
As pointed out in [\dkos], however, there is a more inclusive quantity
which utilizes {\it all} the events and exhibits sensitivity to $\chi(\omega)$.
If we integrate over all photon angles, the total yield can be
written   [\dkos]
\beq
\omega {dN^{(\alpha\beta)} \over d \omega} = {\alpha\over \pi}
\left[ \cR^{(\alpha\beta)}_{\rm ind} +
2 \chi(\omega) \cF^{(\alpha\beta)}(\Theta_{12}) \right] \; ,
\label{total}
\eeq
where $(\alpha\beta) = (ll), (ql), (qq)$. The important point is that the
first,
$\chi$-independent term does not depend on $\Theta_{12}$. The second term is
given by
\beeq
\cF^{(ql)} = \cF^{(qq)} & = & \log\left({\sin\Theta_{12} \over 2}\right) +
 1 \; ,\nonumber \\
\cF^{(ll)} & = & \log\left({1- \cos\Theta_{12} \over 2}\right) + 1 \; .
\eeeq
These can be either positive (smaller $\Theta_{12}$) or negative
(larger $\Theta_{12}$) and the critical angles where the sign change occurs
are readily shown to be [\dkos] $\Theta_{\rm crit} \approx 47.4\degree$ and
$ 74.7\degree$
for the $(ql),(qq)$ and $(ll)$ cases respectively.
If we integrate over events with $\Theta_{12}$ above and below the
critical angle
 and take the {\it difference }
\beeq
\label{deltall}
\delta^{(ll)} &= &
{1\over 1 + \cos\Theta_{\rm crit} }
\int_{\Theta_{\rm crit}}^\pi  \omega {dN^{(ll)}\over d \omega}
\sin\Theta_{12}\; d \Theta_{12}   \nonumber \\
& & \ - \ {1\over 1 - \cos\Theta_{\rm crit} }
\int^{\Theta_{\rm crit}}_0  \omega {dN^{(ll)}\over d \omega}
\sin\Theta_{12}\;  d \Theta_{12}     \; ,  \\
\label{deltaqq}
\delta^{(ql,qq)} &= &
{1\over 1 - \cos\Theta_{\rm crit} }
\int_{\Theta_{\rm crit}}^{\pi/2}  \omega {dN^{(ql,qq)}\over d \omega}
\sin\Theta_{12}\; d \Theta_{12}   \nonumber \\
& & \ - \ {1\over \cos\Theta_{\rm crit} }
\int^{\Theta_{\rm crit}}_0  \omega {dN^{(ql,qq)}\over d \omega}
\sin\Theta_{12}\;  d \Theta_{12}     \; ,
\eeeq
then the $\chi$ independent term in (\ref{total}) cancels and we are left with
\beeq
\delta^{(ll)}  & = &   {\alpha\over \pi}\;  {4\chi(\omega) \over
1 + \cos\Theta_{\rm crit} } \; = \; {\alpha\over \pi}\; 3.164\; \chi(\omega)\;
, \\
 \delta^{(ql)}  =   \delta^{(qq)}  & = &
 {\alpha\over \pi}\;  {2 \chi(\omega) \over \cos\Theta_{\rm crit}
( 1- \cos\Theta_{\rm crit}) } \;
\left[ - \cos\Theta_{\rm crit}  + {1\over 2} \log\left(
{(1+ \cos\Theta_{\rm crit})\over ( 1- \cos\Theta_{\rm crit})} \right) \right]
\nonumber \\
& = &  {\alpha\over \pi}\; 1.343\;  \chi(\omega)\;  .
\eeeq
Fig.~4 shows these quantities as functions of the ratio $\omega/\gw$.
Several remarks are appropriate.
First, the term  $\cR^{(\alpha\beta)}_{\rm ind} $
 in (\ref{total}) contains logarithmic singularities when the photon
is collinear with the (massless) fermions, but these  cancel in the
difference  $\delta^{(\alpha\beta)}$. Second, contributions from
 \lq secondary' photons  in the
quark jets are also expected to cancel in the difference.
 In practice, the
photons can be isolated from the final state jets and leptons
(assuming $\theta_{\rm iso} \ll \Theta_{\rm crit}$) without
significantly weakening the dependence of the  $\delta^{(\alpha\beta)}$
on $\omega/\gw$.

We now return to the question of event rates with which this section
began.  The probablilty of finding
an additional soft photon in a $WW$ event is just given by
the radiation probability given in (2), integrated over photon angles and
energies.
This probability depends on the velocities of the final state particles
(the velocities must be kept to avoid collinear singularities)
as well as the range of photon energies included, but is typically $5 - 10\%$.
  With an integrated luminosity of 500 pb$^{-1}$/yr, LEP200 will
produce about 8000  $W$ pairs each year.  Combining
these numbers with the appropriate branching ratios ($1/81, 12/81,\ {\rm and\ }
36/81$ for the $ll$, $ql$, and $qq$ modes, respectively) implies
event rates ranging from on the order of $10$ (for $ll$) to $100$ (for $qq$)
per year.  Although these
numbers are small, we emphasize again that the asymmetries
$\delta$ defined above make use of {\it all} events while retaining
sensitivity to $\Gamma_W$. However,
many years' running at LEP200 will probably  be required
for the effects described above to be measurable with any precision.
 Future linear colliders with
higher luminosity would certainly fare better.

\section{Concluding Remarks}

A measurement of the total $W$ width, independent of decay modes (and of
the $Z$ width) is by no means easy to obtain.
The method described here, using soft photon radiation,
is limited by statistics; alternatives either
are not direct measurements of the {\it total} width, or have
systematic difficulties of their own.  In any case, a direct measurement
of the $W$ width, at LEP200 or at future colliders if necessary, is
worth pursuing.

In summary, we have shown that soft photon radiation in $WW$ production near
threshold is sensitive to $\gw$ due to width-dependent interference effects.
We have illustrated this sensitivity with some numerical results for angular
distributions for one particular configuration of final state particles.
The configuration we chose was by no means unique, or even optimal, in
its sensitivity. Since soft photons may be a promising way to get at the
$W$ width directly,  more detailed studies of these effects, which take proper
account of event rates and detector capabilities, should
prove worthwhile.

\vskip 1cm
\noindent We are grateful
to P. M\"attig and V.S. Fadin for useful discussions. V.A.K.
thanks the United Kingdom Science and Engineering Research Council for
financial
support.

%\vskip 1cm
\newpage
\noindent{{\Large\bf References}}
\begin{itemize}

\item[{[\tdkbis]}]
Yu.L. Dokshitzer, V.A. Khoze and S.I. Troyan, University of Lund
preprint LU-TP-92-10 (1992).

\item[{[\lepcc]}] See for example: M. Davier {\it et al.},
Proc. ECFA Workshop on LEP 200, Aachen  (1986), eds. A. B\"ohm and
W. Hoogland, report CERN 87-08 (1987), Vol.I, p.120.

\item[{[\dkos]}] Yu.L. Dokshitzer, V.A. Khoze,  L.H. Orr and W.J. Stirling,
University of Durham preprint DTP/92/88 (1992), to be published in Nucl. Phys.
{\bf B} (1993).

\item[{[\kos]}] V.A. Khoze,  L.H. Orr and W.J. Stirling,
Nucl.~Phys. {\bf B378} (1992) 413.

\item[{[\jikia]}]
G. Jikia, Phys. Lett. {\bf 257B} (1991) 196.

\item[{[\veltman]}] M. Lemoine and M. Veltman,
Nucl.~Phys. {\bf B164} (1980) 445.

\item[{[\book]}]
Yu.L. Dokshitzer, V.A. Khoze, A.H. Mueller, and S.I. Troyan,
{\it Basics of Perturbative QCD}, Editions Frontieres, 1991.

\item[{[\jpsi]}] Ya.I. Azimov {\it et al.}, JETP Lett. {\bf 21}
(1975) 378. \\
L.N. Lipatov and V.A. Khoze, Materials of the X Winter School of the Leningrad
Nuclear
Physics Institute, Vol.II, Leningrad (1975) p.409. \\
O.P. Sushkov, Sov. J. Nucl. Phys. {\bf 22} (1975). \\
M. Greco, G. Pancheri-Srivastava and Y. Srivastava,
Nucl. Phys. {\bf B101} (1975) 234. \\
V.N. Baier {\it et al.} Phys. Rep. {\bf 78} (1981) 294.

\end{itemize}
%
%\vskip 1cm
\newpage
\noindent{{\Large\bf Figure Captions}}
\begin{itemize}
\item [{[1]}]
The process $\ee\to\ffff \gamma$ illustrating the angles
used to define the radiation pattern.

\item [{[2]}]
The radiation pattern $\cNql$ as a function of $\theta_0$
for fixed values of $\phi_0$: (a) $\phi_0=0\degree$, (b)
$\phi_0=45\degree$ and  (c) $\phi_0=90\degree$,
  for $\chi=0$ (solid lines) and
$\chi=1$ (dashed lines).

\item [{[3]}]
The radiation pattern $\cNql$ as a function of $\phi_0$
for fixed values of $\theta_0$: (a) $\theta_0=50\degree$, (b)
$\theta_0=75\degree$, (c) $\theta_0=105\degree$ and  (d) $\theta_0=130\degree$,
  for $\chi=0$ (solid lines) and
$\chi=1$ (dashed lines).

\item [{[4]}] The normalized integrated photon yield
differences $\delta^{(\alpha\beta)}$ defined in
Eqs.~(\ref{deltall},\ref{deltaqq}) as functions of $\omega/\gw$, in units of
$\alpha/\pi$.

\end{itemize}

\end{document}